\documentclass[twocolumn,prl,aps,superscriptaddress]{revtex4-1}%
\usepackage{amsmath}
\usepackage{amssymb}
\usepackage{amsfonts}
\usepackage{bm}
\usepackage{amssymb}
\usepackage[dvips]{graphicx}
\usepackage{dcolumn}
\usepackage{txfonts}
\usepackage{makeidx}
\usepackage{color}
\usepackage{mathtools}
\usepackage{threeparttable}
\usepackage[linkcolor=blue,anchorcolor=black,citecolor=blue]{hyperref}

\begin{document}

\title{Room-temperature ultra-sensitive mass spectrometer via dynamic decoupling}
\author{Nan Zhao}
\email{nzhao@csrc.ac.cn; http://www.csrc.ac.cn/~nzhao}
\affiliation{Beijing Computational Science Research Center, Beijing 100084, China}
\author{Zhang-qi Yin}
\affiliation{The Center for Quantum Information, Institute for Interdisciplinary Information Sciences, \\
Tsinghua University, Beijing 100084, P. R. China}
\begin{abstract}
We propose an ultra-sensitive mass spectrometer based on a coupled quantum-bit-oscillator system.
Under dynamical decoupling control of the quantum bit (qubit), the qubit coherence exhibits a comb structure in time domain.
The time-comb structure enables high precision measurement of oscillator frequency, 
which can be used as an ultra-sensitive mass spectrometer.
Surprisingly, in ideal case, the sensitivity of the proposed mass spectrometer, which scales with the temperature $T$ as $T^{-1/2}$, has better performance in higher temperature. 
While taking into account qubit and oscillator decay, we show that the optimal sensitivity is independent on environmental temperature $T$.
With present technology on solid state spin qubit and high-quality optomechanical system, our proposal is feasible to realize an ultra-sensitive mass spectrometer in room temperature. 
\end{abstract}

\maketitle
{\it Introduction-}. 
Single quantum objects, such as single atoms and single photons, attracted more and more attentions in recent years.
Novel applications, such as quantum information processing, triggered fast technique development in {\it isolating} single quantum objects from the noisy environment, 
precisely {\it controlling} their quantum states, 
and {\it hybridizing} different quantum systems. 
The technique development, in turn, provides opportunities of using single quantum objects to design  more distinctive and more powerful tools in various research fields.

Detection of extremely weak signals, such as magnetic fields produced by single nuclei \cite{Zhao2012,Kolkowitz2012a,Taminiau2012a,Mamin2013,Staudacher2013a}, and tiny mass of single molecules \cite{Li2013,Ganzhorn2013,Chaste2012,Jensen2008,Naik2009}, has broad applications in chemistry and biology.
In the past a few decades, detectors based on single quantum objects were designed, and their sensitivity was progressively improved.
For example, using mechanical cantilevers or well-controlled single spins, people are able to detect and resolve single spins of electrons \cite{Rugar2004} and nuclei \cite{Zhao2012,Kolkowitz2012a,Taminiau2012a}. 
For mass sensors, the minimum detectable mass was decreased from femtogram to yoctogram \cite{Ganzhorn2013,Chaste2012,Jensen2008,Naik2009}, reaching the single-proton limit.

In this Letter, we propose an ultra-sensitive measurement scheme based on a coupled quantum-bit-oscillator system. 
We show that, with many-pulse dynamical decoupling (DD) control \cite{Viola1999} on the quantum bit (qubit) of the coupled system, 
the qubit coherence exhibits periodic sharp peaks, forming a {\it comb structure} in time domain. 
The qubit coherence peaks are synchronized with the oscillator period, 
and the peak width decreases when increasing the measurement resource, namely, the DD control pulse number. 
With this time-comb structure of the qubit coherence, tiny changes of the oscillator frequency, 
e.g., due to absorption of a single molecule onto a mechanical oscillator, 
can be monitored and precisely determined from the shift of the coherence peaks.

Two distinctive features make our proposal possible to reach ultra-high sensitivity and have outstanding performance in room temperature.
Firstly, the measurement sensitivity scales with the control pulse number $N$ as $\sim N^{-3/2}$, 
which is different in comparison to $N^{-1/2}-$dependence in the magnetometry schemes using single qubit \cite{DeLange2011}.
The improved scaling relation enables us reach high sensitivity faster, with less measurement resource.
Secondly, we show that the optimal sensitivity is {\it independent} on environmental temperature $T$. 
Furthermore, in high temperature, less control pulses are required to reach the optimal sensitivity and the proposed mass spectrometer has better performance.
For most of the conventional sensing schemes, low-temperature (e.g., liquid Helium temperature) is required, since measurement sensitivity is usually limited by thermal fluctuation which is proportional to $\sqrt{k_{\text{B}}T}$ (with Boltzmann constant $k_{\text{B}}$) \cite{Garbini1995}.
The temperature-independent feature of the optimal sensitivity in our proposal allows novel applications in room temperature, 

\begin{figure}[b]
  \includegraphics[width= \columnwidth]{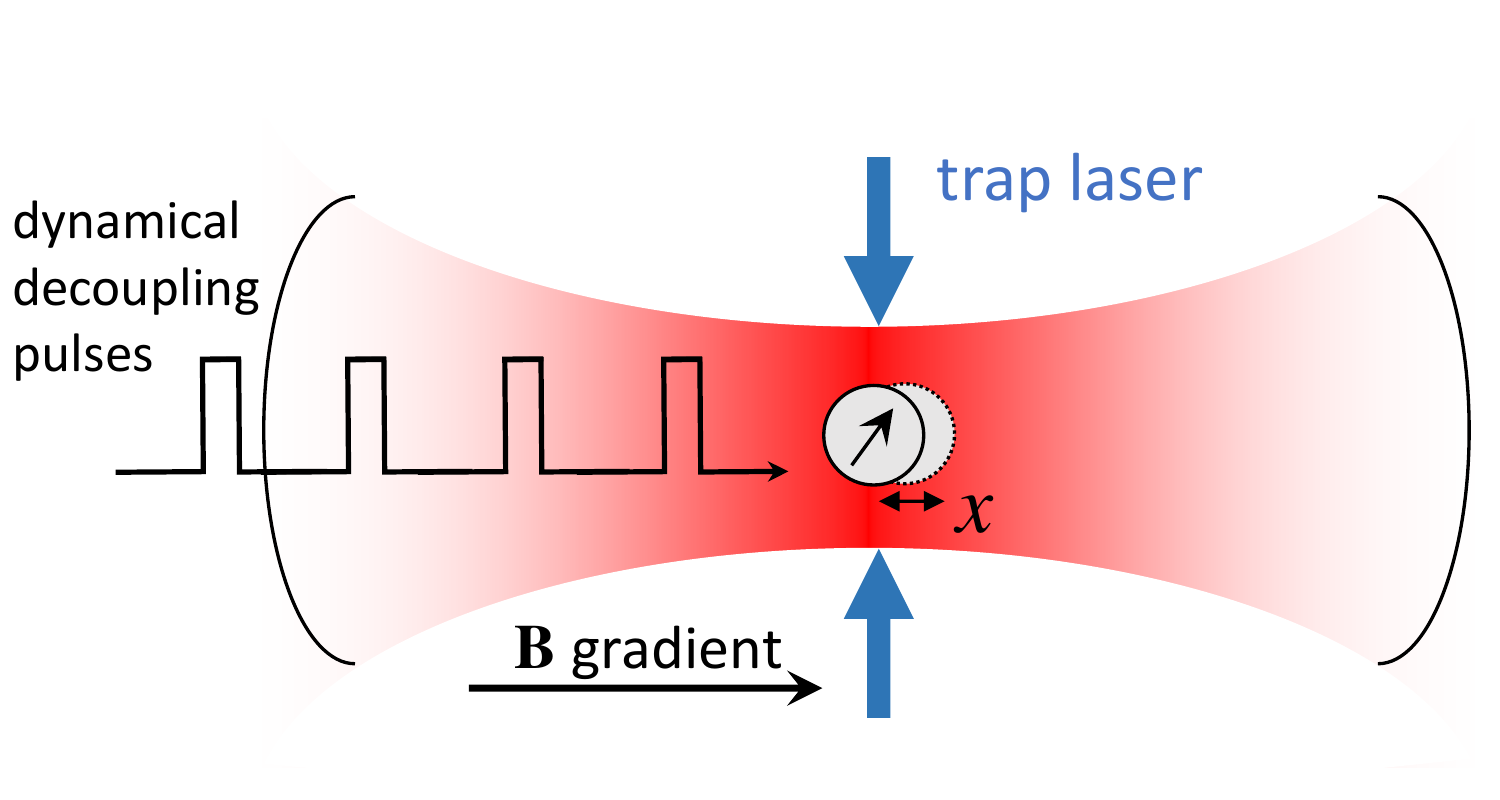}
  \caption{(Color online) Schematic of the proposed mass spectrometer.
  A nano-diamond is trapped in a harmonic potential by counter-propagating laser. 
  The nano-diamond (the grey circle) contains an NV center, which serves as a qubit.
  With a gradient magnetic field, the center-of-mass motion of the nano-diamond couples to the NV center spin.
  Under DD control, the qubit coherence exhibits time-comb structure, 
  which can be used to measure the tiny change of the oscillation frequency (thus the mass change) of the nano-diamond.
  }\label{Fig1}
\end{figure}

Recent technique development on solid-state qubit and mechanical oscillator provides the feasibility of our proposal. 
Solid-state single spin qubit, such as nitrogen-vacancy (NV) centers  \cite{Gruber1997,Wrachtrup2006}, 
has been demonstrated to be well isolated with long coherence time \cite{Balasubramanian2009}.
Meanwhile, the mechanical oscillators of micro- or nano-size have been experimentally fabricated and widely used in detecting weak signals \cite{Garbini1995,Cleland2003}.
Particularly, the recent optomechanical systems \cite{Marquardt2009,Aspelmeyer2013,Li2013}, optically levitated particles \cite{Romero-Isart2010,Chang2010,Li2010a,Li2011}, 
are believed to reach high quality factor up to $10^{10}$ \cite{Yin2013a} or even higher, 
which enables such systems to detect novel quantum effect \cite{Geraci2010,Arvanitaki2013,Yin2011}.
Here, we combine the advanced qubit and opomechanical systems, and propose that hybrid systems such as optically levitated nano-diamond with a single NV center \cite{Neukirch2013, Scala2013,Yin2013} can realize a high mass sensitivity up to of $10^{-22}\text{g}/\sqrt{\text{Hz}}$ in room temperature.

{\it Time-comb under dynamical decoupling-}. We consider a coupled qubit-oscillator model with the Hamiltonian \cite{Scala2013, Yin2013, Kolkowitz2012}
\begin{equation}
H=\frac{1}{2}\omega_{\text{q}}\sigma_z + \omega_0 b^{\dagger}b + \frac{1}{2} \lambda \sigma_z(b^{\dagger}+b),
\label{harmonic Hamiltonian}
\end{equation}
where $\omega_0$ ($\omega_\text{q}$) is the frequency of the harmonic oscillator (qubit), 
and $\lambda$ is the coupling strength.
The qubit is initially prepared in a superposition state 
$\vert\psi(0)\rangle_{\text{q}} = \left(\vert 0\rangle + \vert 1\rangle\right)/\sqrt{2}$.
The oscillator is initially in a thermal equilibrium state  $\rho_{\text{b}} = Z^{-1}\exp\left(-\beta \omega_0 b^{\dagger}b\right)$ with the partition function 
$Z=\text{Tr}\left[\exp\left(-\beta \omega_0 b^{\dagger}b\right)\right]$
and the inverse environmental temperature $\beta=1/(k_{\text{B}}T)$.

\begin{figure}[tb]
  \includegraphics[width= \columnwidth]{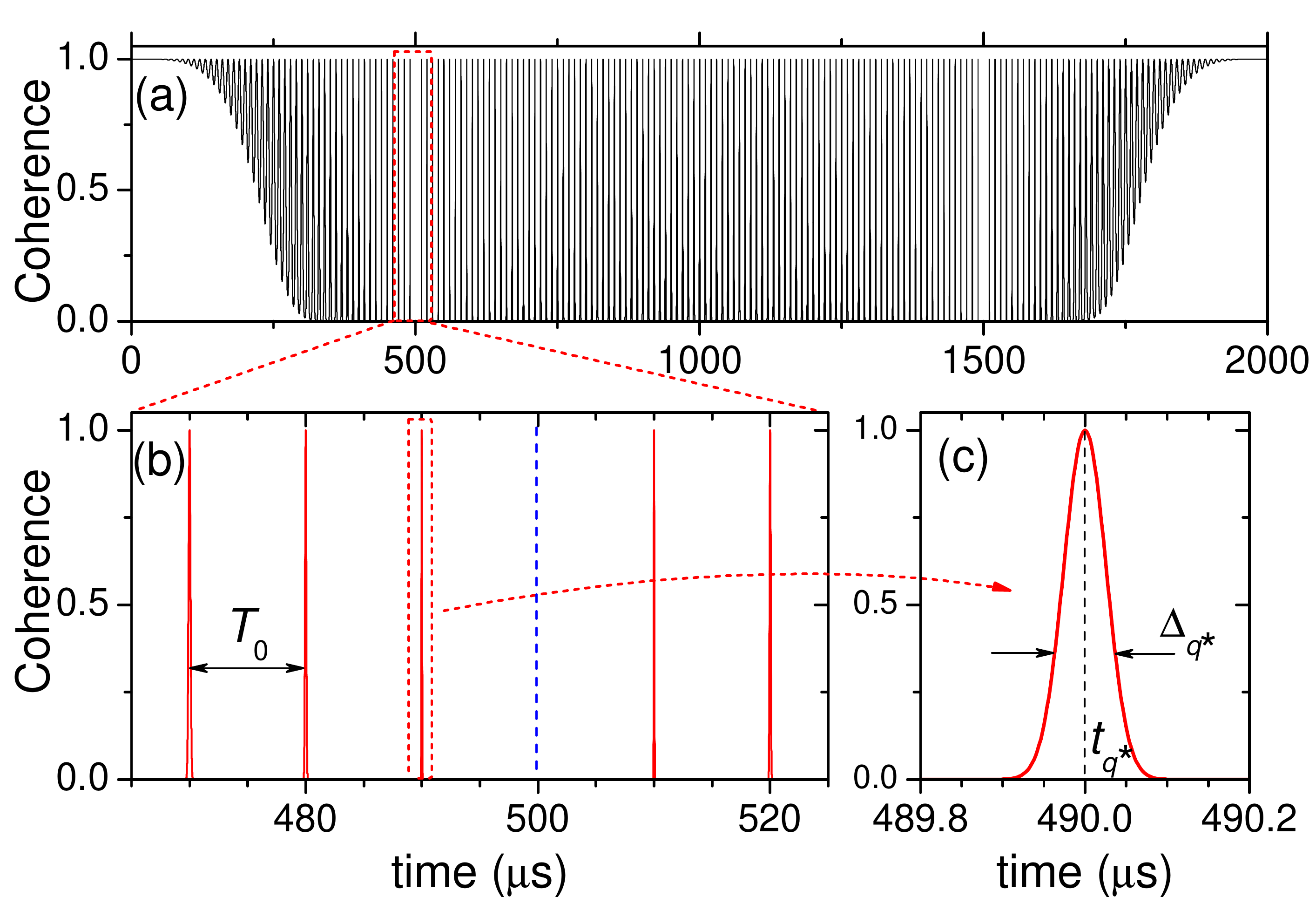}
  \caption{(Color online) (a) The time-comb structure of qubit coherence under 100-pulse CPMG control. 
  For $\omega_0 t\gg 1$, the comb period is synchronized with the oscillator period $T_0=2\pi/\omega_0$.
  (b) Close-up of the coherence peaks. A missing peak at $\omega_0 t = N \pi$ is indicated by the blue dashed line. 
  The peak width is decreasing when getting close to the missing one.
  (c) Close-up of the narrowest coherence peak,  which is centered at $t_{q^{*}}$ with width $\Delta_{q^{*}}$ (see text).
  The parameters used in this figure are oscillator frequency $\omega_0/(2\pi)=100$~kHz, coupling strength $\lambda=0.001\omega_0$, temperature $T=10$~K and 100-pulse CPMG control.
  }\label{Fig2}
\end{figure}

Since $\sigma_z$ is a good quantum number in Eq.~(\ref{harmonic Hamiltonian}), we focus on the dynamics of the relative phase, or quantum coherence \cite{Yang2010c}, between qubit states $\vert 0\rangle$ and $\vert 1\rangle$ influenced by the oscillator.
Under DD control of the qubit, which flips the qubit state by a train of $\pi$ pulses applied at times $t_j$ for $j=1,2,\dots, N$ (with $t_0\equiv 0$ and $t_{N+1} \equiv t$), 
the qubit coherence $L(t)$ is expressed as \cite{Yang2008}
\begin{equation}
L(t)=\left\langle \mathcal{T}_c e^{-i\int_c \hat{X}(t^{\prime}) f(t^{\prime}) dt^{\prime}} \right\rangle,
\label{coherence}
\end{equation}
where the integral is performed on a time contour $c:0\rightarrow t\rightarrow 0$, $\mathcal{T}_c$ is the contour-time-ordering operator, and $\hat{X}(t)=\lambda \left(b^{\dagger}e^{i\omega_0 t}+b e^{i\omega_0 t}\right)$ is proportional to the oscillator displacement in the interaction picture. 
The qubit flipping by DD control is described by the sign function $f(t)$, which toggles between $+1$ or $-1$ whenever a $\pi$-pulse is applied.

The Gaussian statistics nature of the harmonic oscillator allows the coherence in Eq.~(\ref{coherence}) to be exactly evaluated  \cite{Zhao2011a, Kolkowitz2012}.
Particularly, under the $N$-pulse CPMG sequence [with $N$ qubit flips at $t_j = (2j-1)/(2N)$], the qubit coherence is  $L(t) = \exp\left[-\chi(t)/2\right]$ with \cite{Zhao2011a, Kolkowitz2012}
\begin{eqnarray}
\chi(t) &=&\int_0^{\infty} \frac{d\omega}{\pi} S(\omega)\left\vert F(\omega t)\right\vert^2 =\frac{4\tilde{\lambda}^2}{\omega_0^2}  \left(\sec\frac{\omega_0 t}{2N}-1\right)^2\sin^{2}\frac{\omega_0 t}{2}\notag\\
&\equiv& \Gamma^2(t)\sin^2\frac{\omega_0 t}{2},
\label{CPMG chi}
\end{eqnarray}
where $S(\omega)=\tilde{\lambda}^2\pi \delta(\omega-\omega_0)$ is the noise spectrum of the oscillator, $F(\omega t)$ is the Fourier transform of the modulation function $f(t)$, $\tilde{\lambda}^2=\lambda^2(2n_{\text{th}}+1)$, and $n_{\text{th}}\equiv \left[\exp(\beta \omega_0)-1\right]^{-1}$ is the  thermal occupation number of the oscillator. In the second line of Eq~(\ref{CPMG chi}), $\Gamma(t)$ is a slow-varying envelope function. 

The qubit coherence exhibits novel dynamics with many-pulse ($N\gg1$) DD, as shown in Fig.~\ref{Fig2}.
In short time limit ($\omega_0 t\ll N\pi$), the qubit coherence is well-protected (close to unity) by the DD control. 
As increasing time $t$, the qubit coherence become oscillating. 
Furthermore, when $\Gamma(t)\gg 1$, the qubit enters a new regime where the coherence almost decays completely [$L(t)\approx 0$], 
{\it except} in the narrow intervals around the zero points of Eq.~(\ref{CPMG chi}), i.e., $t_q=qT_0$ [for integer $q$ and $q\neq (2k+1)N/2$], where $T_0=2\pi/\omega_0$ is the oscillator period.
In this regime, the qubit coherence forms a {\it comb structure}.

In the time-comb regime, the coherence shows sharp peaks in Gaussian shapes $L(t)\approx  e^{- \gamma^2_q\left(t-t_q\right)^2/2}$ [see. Fig.~\ref{Fig2}(c)].
The peak width decreases when $t_q$ approaches odd-multiple of $N \pi/\omega_0$ [i.e. the diverge point of $\Gamma(t)$, indicated by the blue dashed line in Fig.~\ref{Fig2}(b)].
For a given control pulse number $N$, the narrowest peak (for $q= q^*\equiv N/2-1$) appears at $t_{q^{*}}$ and with peak width $\Delta_{q^{*}}$
\begin{eqnarray}
t_{q^{*}}&=&\left(\frac{N}{2}-1\right)T_0,\notag\\
 \Delta_{{q}^{*}}&\equiv& \frac{2\sqrt{2}}{\gamma_{q^*}} =\frac{T_0}{N\Lambda\sqrt{2n_{\text{th}}+1}},
\label{tQ expression}
\end{eqnarray}
where $\Lambda=\lambda/\omega_0$ is the ratio of the coupling strength to the oscillator frequency. 
Notice that peak width is {\it inversely proportional} to the control pulse number $N$ and the square root of the thermal excitation number $n_{\text{th}}$ (for $n_{\text{th}}\gg1$).
In the following, we will show that the peak narrowing with increasing $N$ or $n_{\text{th}}$ (i.e., increasing temperature $T$) improves the sensitivity.

{\it Ultra-sensitive mass spectrometer.}-
We propose that the time-comb can be used to measure tiny mass change of the mechanical oscillator, 
e.g. due to absorption of single molecules, 
by monitoring shifts of the qubit coherence peaks. 
For an oscillator with oscillation frequency $\omega_0 = \sqrt{k/M}$ 
(for $k$ and $M$ being the spring constant and the mass, respectively), 
a small change $\delta M$ of the mass $M$ induces a change $\delta L$ of the coherence $L(t)$ around the recovery peak. 
The relative mass uncertainty is
\begin{equation}
\frac{\delta M}{M}  = \frac{2\delta \omega_0}{\omega_0} =  \frac{1}{\gamma_{q^*}^2(t-t_{q^{*}})t_{q^{*}}}\frac{\delta L}{L} \approx \frac{1}{\gamma_{q^*} t_{q^{*}}} \frac{\delta L}{L},
\end{equation}
where we have chosen a proper measurement time $t$ close to the peak time $t_{q^{*}}$ so that $\gamma_{q^*} (t-t_{q^{*}}) \approx 1$.

We consider the case where the qubit coherence is obtained by averaging the output of $N_{\text{run}}$ independent Bernoulli trials. 
In this case, the uncertainty $\delta L$ comes from the shot-noise in the measurement, i.e. $\delta L/L \approx N_{\text{run}}^{-1/2}$. For the total measurement time $T_{\text{tot}}= N_{\text{run}} t_{q^{*}}$,
the mass sensitivity $\eta_M$, up to a constant of the order of unity, is
\begin{equation}
\eta_M\equiv \delta M \sqrt{T_{\text{tot}}}= \frac{M}{(2N)^{3/2}\Lambda  \sqrt{k_{\text{B}}T/h}},
\label{IdealSensitivity}
\end{equation}
where $h$ is Planck constant. Here we have assumed that $N\gg 1$ and $n_{\text{th}}\approx k_{\text{B}}T/\hbar \omega_0\gg 1$, which is the case for most practical mechanical oscillator systems.

\begin{figure}[tb]
  \includegraphics[width= \columnwidth]{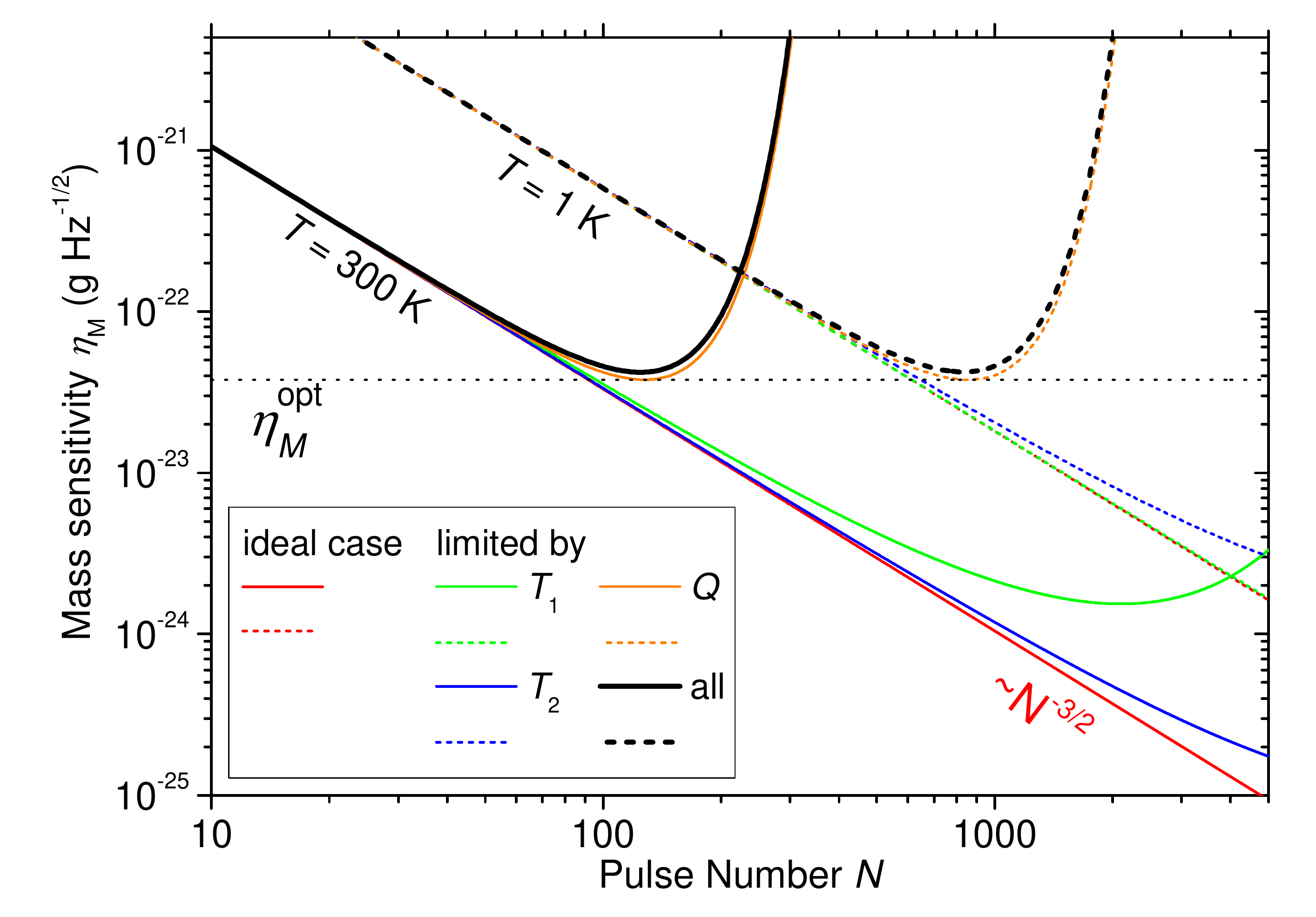}
  \caption{(Color online) Mass sensitivity $\eta_M$ as functions of DD control pulse number $N$.
  Solid curves are the sensitivity in room temperature $T=300$~K, while dashed curves are the sensitivity in low temperature $T=1$~K.
  The red curves are the ideal sensitivity according to Eq.~(\ref{IdealSensitivity}), which scales as $\sim N^{-3/2}$.
  The curves in green, blue, and orange are the sensitivity taking into account the qubit $T_1$-decay (with $T_1=7$~ms), qubit $T_2$-decay (with $T_2=100~\mu\text{s}$), and oscillator finite $Q$-factor (with $Q=10^9$), in turn.
  The thick black curves are the sensitivity with all the decay mechanisms included.
  The horizontal dotted line indicates the temperature-independent optimal sensitivity.
  Other parameters ($\omega_0$ and $\lambda$) used in this figure are the same as those in Fig.~\ref{Fig2}.
  }\label{Fig3}
\end{figure}

Equation~(\ref{IdealSensitivity}) reveals two interesting features of the qubit-oscillator based mass spectrometer. Firstly, the scaling relation of sensitivity to the pulse number $N$ is different with that appears in magnetometry. 
In the case of using qubit for magnetometry under DD, the sensitivity scales with the control pulse number as $\sim N^{-1/2}$  \cite{DeLange2011}. In our case, the narrowing effect of the peak when increasing the pulse number $N$ improves the scaling relation to $\sim N^{-3/2}$, 
which will help to achieve the optimal sensitivity faster.

Secondly, and more interestingly, the sensitivity is \textit{inverse-linearly} dependent on the square root of temperature. 
For the conventional oscillator-based sensor, the sensitivity is usually limited by thermal fluctuation of the oscillator displacement $x$, which is characterized by the root-mean-square amplitude $x_{\text{rms}}=\sqrt{k_{\text{B}}T/k}$. 
High temperature would destroy the sensitivity, and prevent the room-temperature applications.
While in our measurement scheme, the measured quantity $\omega_0$ does not directly couple to the oscillator displacement $x$ and, thus, its uncertainty is independent on the position thermal fluctuation.
Instead, the more thermal phonons in higher temperature cause stronger effective coupling between the qubit and oscillator, which improves the sensitivity.

{\it Sensitivity limitations.}-
Now we analyse the factors which limit the ideal sensitivity shown in Eq.~(\ref{IdealSensitivity}).
The qubit decoherence (including relaxation and dephasing) and the oscillation dissipation caused by the inevitable coupling to the environment are the two reasons which set lower bound to the sensitivity.

The environmental fluctuation to the qubit, which causes the qubit decoherence, prevents the perfect recovery of coherence shown in Fig.~\ref{Fig2}. With both longitudinal relaxation process (or $T_1$-process) and transverse relaxation process (or $T_2$-process), the qubit suffers a background decoherence $L_{\text{bg}}(t)$ in additional to the oscillator-induced periodic revival peaks. 
Taking the solid state spin qubit for example, the background decoherence can be modelled as  $L_{\text{bg}}(t)= \exp\left[-t/T_1 - (t/T_2^{(N)})^3\right]$  \cite{Lange2010}. 
The longitudinal decoherence, typically caused by phonon scattering, is a Markovian proceess (a simple exponential decay), and is hardly corrected by DD.
The the transverse decoherence, usually caused by spin baths, can be protected by DD with the decay time $T_2^{(N)}$ depending on the DD control pulse number $N$ as $T_2^{(N)}=T_2 N^{2/3}$  \cite{Lange2010} (for $T_2$ being the coherence time for $N=1$). 

The background qubit decoherence $L_{\text{bg}}(t)$ reduces the height of the recover peaks. 
Consequently, the mass sensitivity is magnified by a factor of $L^{-1}_{\text{bg}}(t_{q^{*}})$.
The balance between the $\sim N^{-3/2}$ sensitivity scaling and the background decoherence gives rise to an optimal pulse number to the sensitivity (see Fig.~\ref{Fig3}), similar to Ref.~\cite{DeLange2011} but with an additional peak narrowing effect. 
Qubit with long coherence time, like NV centers in diamond, can be chosen to eliminate the background decoherence $L_{\text{bg}}(t)$ effect.
In low temperature, the $T_1$ time can reach the order of seconds \cite{Takahashi2008}, and the $T_2$ time has been demonstrated to be $\sim \text{ms}$ or even longer under DD control \cite{Bar-Gill2011}.
In this case, the qubit decoherence becomes less important, and the dissipation of mechanical oscillation is the dominant mechanism limiting the sensitivity.

The coupling to the environment of mechanical resonator causes broadening of the oscillator frequency.
In Eq.~(\ref{CPMG chi}), with the $\delta$-function in the noise spectrum $S(\omega)$ replaced by a Lorentzian spectrum with finite broadening $\kappa = \omega_0/Q$ [i.e. $S(\omega) = \tilde{\lambda}^2\kappa/[(\omega - \omega_0)^2 +\kappa^2]$, for $Q$ being the quality factor], 
the coherence cannot recover perfectly even though the central oscillation frequency hits the zero points of the filter function. 
The overlap between the wings of the Lorentzian spectrum and the non-zero region of the function $F(\omega t)$ causes the reduction of the hight of the coherence recovery peak [see Eq.~(\ref{CPMG chi})]. 
In the case of $Q\gg N\gg 1$, the function $\chi(t)$ around the recovery time $t=t_{q^{*}}$ is calculated as
$\chi(t)\approx \chi(t_{q^{*}}) + \gamma_{q^*}^2 (t-t_{q^{*}} )^2$ with $\chi(t_{q^{*}})=4\tilde{\lambda}^2 N^3 /(\omega_0^2 Q)$.
Increasing the pulse number $N$ reduces the recovery height.
The optimal control pulse number $N_{\text{opt}}$ is estimated by $\chi(t_{q^{*}})\approx 1$, which gives 
\begin{equation}
N_{\text{opt}}\approx\frac{\omega_0}{2\lambda}\left(\frac{\hbar \lambda Q}{ k_{\text{B}}T}\right)^{\frac{1}{3}}.
\label{OptimalPulseNum}
\end{equation}
Substitute the optimal pulse number $N_{\text{opt}}$ into Eq.~(\ref{IdealSensitivity}), one obtains a universal value of the optimal mass sensitivity (up to a constant of the order of unity)
\begin{equation}
\eta_M^{\text{opt}} = \frac{M}{\sqrt{f_0 Q}},
\label{OptimalSenisitivity}
\end{equation}
where $f_0=\omega_0/(2\pi)$.
In the this case, the optimal sensitivity only relies on the properties of the oscillator (i.e., $f_0$ and $Q$), 
and is {\it independent} on the temperature $T$ nor the qubit-oscillator coupling strength $\lambda$.
In addition to the temperature-independent feature, 
we notice that in higher temperature, less controlled pulses are required to reach the optimal sensitivity according to Eq.~(\ref{OptimalPulseNum}) (see Fig.~\ref{Fig3}).
In this sense, the proposed mass spectrometer has better performance in high themperature, in sharp contrast to conventional schemes that low temperature is necessary to reduce the thermal fluctuation.
The universal form of the optimal sensitivity in Eq.~(\ref{OptimalSenisitivity}) provides a simple guiding principle to design the system and to experimentally implement our proposal.
The model and the sensitivity described in Eqs.~(\ref{harmonic Hamiltonian})-(\ref{OptimalSenisitivity}), indeed, is quite general, and can be realized in different types of system. 
In the following, we take the optically levitated nano-diamond with NV center as an example to demonstrate the application.

{\it Experimental feasibility.-}
As we discussed above, solid state spin with long coherence time, like nitrogen-vacancy in diamond, serves as a good candidate of the qubit. The $T_1$ and $T_2$ decoherence has negligible effect in the sensitivity (see Fig.~\ref{Fig3}). 
Meanwhile, the coherent coupling between NV center electron spin and the mechanical motion has been demonstrated very recently \cite{Kolkowitz2012} with a low-$Q$ mechanical cantilever.
Since high quality factor $Q$ of the mechanical oscillator is essential for the ultimate sensitivity,
we propose that the system of optically levitated nano-diamond with a single NV center is a good candidate for realizing the ultra-sensitive mass spectrometer,
where the quality factor can reach $\sim 10^{10}$ or even higher.

Consider a nano-diamond of 50~nm in diameter (corresponding to the mass $M=2.3\times10^{-16}$~g) which is optically trapped in a harmonic potential with center-of-mass (CoM) oscillating frequency $f_0= 100~\text{kHz}$. In a gradient magnetic field, the CoM motion of the nano-diamond couples to the NV center electron spin in the nano-diamond in a manner described in Eq.~(\ref{harmonic Hamiltonian}).
With the magnetic field gradient $G_m=200~\text{T/m}$, the coupling strength is $\sim 100~\text{Hz}$.
With the quality factor $Q=10^9$, the system can reach a mass sensitivity of the order $\sim 10^{-22}~\text{g}/\sqrt{\text{Hz}}$. 
In practical experiments, the efficiency of the optical readout of the spin state of NV center is limited by the spin-selective fluorescence
contrast, and the photon collection efficiency. 
This gives rise to a technique factor $1/C \approx 10^{-2} \sim  10^{-1}$ unfavourable for the sensitivity  \cite{Taylor2008}. 
However, even though the technique factor may deteriorate the sensitivity by one or two orders of magnitudes,
the temperature-independent feature of our proposed system will be still attractive for a room temperature sensor.

{\it Conclusion.-}
In this paper, we propose an ultra-sensitive mass spectrometer based on the coupled qubit-oscillator system.
Using the many-pulse DD technique, the qubit coherence exhibits a time-comb structure, which enable the precise measurement of the oscillating frequency. 
The combination of advanced techniques on NV center in diamond and optical levitated nano-particle, which serve as long-live qubit and high quality oscillator respectively, makes the room-temperature ultra-sensitive mass spectrometer ready to be realized.

We thank Dr. L. Ge, Professor Y. Li, Professor W. Yang, Professor C.P. Sun, and Professor J. Twamley for fruitful discussion about theoretical scheme, 
and Dr. T.C. Li, Dr. F. Xue, Dr. Y. Tao for suggestions on the experimental feasibility.
N.Z. is supported by NSFC No. 11374032. Z.Q.Y. is funded by the NBRPC (973 Program) 2011CBA00300 (2011CBA00302), NNSFC 11105136, and 61033001. 

\bibliographystyle{apsrev4-1}

\begin{thebibliography}{39}%
\makeatletter
\providecommand \@ifxundefined [1]{%
 \@ifx{#1\undefined}
}%
\providecommand \@ifnum [1]{%
 \ifnum #1\expandafter \@firstoftwo
 \else \expandafter \@secondoftwo
 \fi
}%
\providecommand \@ifx [1]{%
 \ifx #1\expandafter \@firstoftwo
 \else \expandafter \@secondoftwo
 \fi
}%
\providecommand \natexlab [1]{#1}%
\providecommand \enquote  [1]{``#1''}%
\providecommand \bibnamefont  [1]{#1}%
\providecommand \bibfnamefont [1]{#1}%
\providecommand \citenamefont [1]{#1}%
\providecommand \href@noop [0]{\@secondoftwo}%
\providecommand \href [0]{\begingroup \@sanitize@url \@href}%
\providecommand \@href[1]{\@@startlink{#1}\@@href}%
\providecommand \@@href[1]{\endgroup#1\@@endlink}%
\providecommand \@sanitize@url [0]{\catcode `\\12\catcode `\$12\catcode
  `\&12\catcode `\#12\catcode `\^12\catcode `\_12\catcode `\%12\relax}%
\providecommand \@@startlink[1]{}%
\providecommand \@@endlink[0]{}%
\providecommand \url  [0]{\begingroup\@sanitize@url \@url }%
\providecommand \@url [1]{\endgroup\@href {#1}{\urlprefix }}%
\providecommand \urlprefix  [0]{URL }%
\providecommand \Eprint [0]{\href }%
\providecommand \doibase [0]{http://dx.doi.org/}%
\providecommand \selectlanguage [0]{\@gobble}%
\providecommand \bibinfo  [0]{\@secondoftwo}%
\providecommand \bibfield  [0]{\@secondoftwo}%
\providecommand \translation [1]{[#1]}%
\providecommand \BibitemOpen [0]{}%
\providecommand \bibitemStop [0]{}%
\providecommand \bibitemNoStop [0]{.\EOS\space}%
\providecommand \EOS [0]{\spacefactor3000\relax}%
\providecommand \BibitemShut  [1]{\csname bibitem#1\endcsname}%
\let\auto@bib@innerbib\@empty
\bibitem [{\citenamefont {Zhao}\ \emph {et~al.}(2012)\citenamefont {Zhao},
  \citenamefont {Ho},\ and\ \citenamefont {Liu}}]{Zhao2012}%
  \BibitemOpen
  \bibfield  {author} {\bibinfo {author} {\bibfnamefont {N.}~\bibnamefont
  {Zhao}}, \bibinfo {author} {\bibfnamefont {S.-w.}\ \bibnamefont {Ho}}, \ and\
  \bibinfo {author} {\bibfnamefont {R.-b.}\ \bibnamefont {Liu}},\ }\href
  {\doibase 10.1103/PhysRevB.85.115303} {\bibfield  {journal} {\bibinfo
  {journal} {Physical Review B}\ }\textbf {\bibinfo {volume} {85}},\ \bibinfo
  {pages} {115303} (\bibinfo {year} {2012})}\BibitemShut {NoStop}%
\bibitem [{\citenamefont {Kolkowitz}\ \emph
  {et~al.}(2012{\natexlab{a}})\citenamefont {Kolkowitz}, \citenamefont
  {Unterreithmeier}, \citenamefont {Bennett},\ and\ \citenamefont
  {Lukin}}]{Kolkowitz2012a}%
  \BibitemOpen
  \bibfield  {author} {\bibinfo {author} {\bibfnamefont {S.}~\bibnamefont
  {Kolkowitz}}, \bibinfo {author} {\bibfnamefont {Q.~P.}\ \bibnamefont
  {Unterreithmeier}}, \bibinfo {author} {\bibfnamefont {S.~D.}\ \bibnamefont
  {Bennett}}, \ and\ \bibinfo {author} {\bibfnamefont {M.~D.}\ \bibnamefont
  {Lukin}},\ }\href {\doibase 10.1103/PhysRevLett.109.137601} {\bibfield
  {journal} {\bibinfo  {journal} {Physical Review Letters}\ }\textbf {\bibinfo
  {volume} {109}},\ \bibinfo {pages} {137601} (\bibinfo {year}
  {2012}{\natexlab{a}})}\BibitemShut {NoStop}%
\bibitem [{\citenamefont {Taminiau}\ \emph {et~al.}(2012)\citenamefont
  {Taminiau}, \citenamefont {Wagenaar}, \citenamefont {van~der Sar},
  \citenamefont {Jelezko}, \citenamefont {Dobrovitski},\ and\ \citenamefont
  {Hanson}}]{Taminiau2012a}%
  \BibitemOpen
  \bibfield  {author} {\bibinfo {author} {\bibfnamefont {T.~H.}\ \bibnamefont
  {Taminiau}}, \bibinfo {author} {\bibfnamefont {J.~J.~T.}\ \bibnamefont
  {Wagenaar}}, \bibinfo {author} {\bibfnamefont {T.}~\bibnamefont {van~der
  Sar}}, \bibinfo {author} {\bibfnamefont {F.}~\bibnamefont {Jelezko}},
  \bibinfo {author} {\bibfnamefont {V.~V.}\ \bibnamefont {Dobrovitski}}, \ and\
  \bibinfo {author} {\bibfnamefont {R.}~\bibnamefont {Hanson}},\ }\href
  {\doibase 10.1103/PhysRevLett.109.137602} {\bibfield  {journal} {\bibinfo
  {journal} {Physical Review Letters}\ }\textbf {\bibinfo {volume} {109}},\
  \bibinfo {pages} {137602} (\bibinfo {year} {2012})}\BibitemShut {NoStop}%
\bibitem [{\citenamefont {Mamin}\ \emph {et~al.}(2013)\citenamefont {Mamin},
  \citenamefont {Kim}, \citenamefont {Sherwood}, \citenamefont {Rettner},
  \citenamefont {Ohno}, \citenamefont {Awschalom},\ and\ \citenamefont
  {Rugar}}]{Mamin2013}%
  \BibitemOpen
  \bibfield  {author} {\bibinfo {author} {\bibfnamefont {H.~J.}\ \bibnamefont
  {Mamin}}, \bibinfo {author} {\bibfnamefont {M.}~\bibnamefont {Kim}}, \bibinfo
  {author} {\bibfnamefont {M.~H.}\ \bibnamefont {Sherwood}}, \bibinfo {author}
  {\bibfnamefont {C.~T.}\ \bibnamefont {Rettner}}, \bibinfo {author}
  {\bibfnamefont {K.}~\bibnamefont {Ohno}}, \bibinfo {author} {\bibfnamefont
  {D.~D.}\ \bibnamefont {Awschalom}}, \ and\ \bibinfo {author} {\bibfnamefont
  {D.}~\bibnamefont {Rugar}},\ }\href {\doibase 10.1126/science.1231540}
  {\bibfield  {journal} {\bibinfo  {journal} {Science (New York, N.Y.)}\
  }\textbf {\bibinfo {volume} {339}},\ \bibinfo {pages} {557} (\bibinfo {year}
  {2013})}\BibitemShut {NoStop}%
\bibitem [{\citenamefont {Staudacher}\ \emph {et~al.}(2013)\citenamefont
  {Staudacher}, \citenamefont {Shi}, \citenamefont {Pezzagna}, \citenamefont
  {Meijer}, \citenamefont {Du}, \citenamefont {Meriles}, \citenamefont
  {Reinhard},\ and\ \citenamefont {Wrachtrup}}]{Staudacher2013a}%
  \BibitemOpen
  \bibfield  {author} {\bibinfo {author} {\bibfnamefont {T.}~\bibnamefont
  {Staudacher}}, \bibinfo {author} {\bibfnamefont {F.}~\bibnamefont {Shi}},
  \bibinfo {author} {\bibfnamefont {S.}~\bibnamefont {Pezzagna}}, \bibinfo
  {author} {\bibfnamefont {J.}~\bibnamefont {Meijer}}, \bibinfo {author}
  {\bibfnamefont {J.}~\bibnamefont {Du}}, \bibinfo {author} {\bibfnamefont
  {C.~a.}\ \bibnamefont {Meriles}}, \bibinfo {author} {\bibfnamefont
  {F.}~\bibnamefont {Reinhard}}, \ and\ \bibinfo {author} {\bibfnamefont
  {J.}~\bibnamefont {Wrachtrup}},\ }\href {\doibase 10.1126/science.1231675}
  {\bibfield  {journal} {\bibinfo  {journal} {Science (New York, N.Y.)}\
  }\textbf {\bibinfo {volume} {339}},\ \bibinfo {pages} {561} (\bibinfo {year}
  {2013})}\BibitemShut {NoStop}%
\bibitem [{\citenamefont {Li}\ and\ \citenamefont {Zhu}(2013)}]{Li2013}%
  \BibitemOpen
  \bibfield  {author} {\bibinfo {author} {\bibfnamefont {J.-J.}\ \bibnamefont
  {Li}}\ and\ \bibinfo {author} {\bibfnamefont {K.-D.}\ \bibnamefont {Zhu}},\
  }\href {\doibase 10.1016/j.physrep.2012.11.003} {\bibfield  {journal}
  {\bibinfo  {journal} {Physics Reports}\ }\textbf {\bibinfo {volume} {525}},\
  \bibinfo {pages} {223} (\bibinfo {year} {2013})}\BibitemShut {NoStop}%
\bibitem [{\citenamefont {Ganzhorn}\ \emph {et~al.}(2013)\citenamefont
  {Ganzhorn}, \citenamefont {Klyatskaya}, \citenamefont {Ruben},\ and\
  \citenamefont {Wernsdorfer}}]{Ganzhorn2013}%
  \BibitemOpen
  \bibfield  {author} {\bibinfo {author} {\bibfnamefont {M.}~\bibnamefont
  {Ganzhorn}}, \bibinfo {author} {\bibfnamefont {S.}~\bibnamefont
  {Klyatskaya}}, \bibinfo {author} {\bibfnamefont {M.}~\bibnamefont {Ruben}}, \
  and\ \bibinfo {author} {\bibfnamefont {W.}~\bibnamefont {Wernsdorfer}},\
  }\href {\doibase 10.1038/nnano.2012.258} {\bibfield  {journal} {\bibinfo
  {journal} {Nature nanotechnology}\ }\textbf {\bibinfo {volume} {8}},\
  \bibinfo {pages} {165} (\bibinfo {year} {2013})}\BibitemShut {NoStop}%
\bibitem [{\citenamefont {Chaste}\ \emph {et~al.}(2012)\citenamefont {Chaste},
  \citenamefont {Eichler}, \citenamefont {Moser}, \citenamefont {Ceballos},
  \citenamefont {Rurali},\ and\ \citenamefont {Bachtold}}]{Chaste2012}%
  \BibitemOpen
  \bibfield  {author} {\bibinfo {author} {\bibfnamefont {J.}~\bibnamefont
  {Chaste}}, \bibinfo {author} {\bibfnamefont {a.}~\bibnamefont {Eichler}},
  \bibinfo {author} {\bibfnamefont {J.}~\bibnamefont {Moser}}, \bibinfo
  {author} {\bibfnamefont {G.}~\bibnamefont {Ceballos}}, \bibinfo {author}
  {\bibfnamefont {R.}~\bibnamefont {Rurali}}, \ and\ \bibinfo {author}
  {\bibfnamefont {a.}~\bibnamefont {Bachtold}},\ }\href {\doibase
  10.1038/nnano.2012.42} {\bibfield  {journal} {\bibinfo  {journal} {Nature
  nanotechnology}\ }\textbf {\bibinfo {volume} {7}},\ \bibinfo {pages} {301}
  (\bibinfo {year} {2012})}\BibitemShut {NoStop}%
\bibitem [{\citenamefont {Jensen}\ \emph {et~al.}(2008)\citenamefont {Jensen},
  \citenamefont {Kim},\ and\ \citenamefont {Zettl}}]{Jensen2008}%
  \BibitemOpen
  \bibfield  {author} {\bibinfo {author} {\bibfnamefont {K.}~\bibnamefont
  {Jensen}}, \bibinfo {author} {\bibfnamefont {K.}~\bibnamefont {Kim}}, \ and\
  \bibinfo {author} {\bibfnamefont {a.}~\bibnamefont {Zettl}},\ }\href
  {\doibase 10.1038/nnano.2008.200} {\bibfield  {journal} {\bibinfo  {journal}
  {Nature nanotechnology}\ }\textbf {\bibinfo {volume} {3}},\ \bibinfo {pages}
  {533} (\bibinfo {year} {2008})}\BibitemShut {NoStop}%
\bibitem [{\citenamefont {Naik}\ \emph {et~al.}(2009)\citenamefont {Naik},
  \citenamefont {Hanay}, \citenamefont {Hiebert}, \citenamefont {Feng},\ and\
  \citenamefont {Roukes}}]{Naik2009}%
  \BibitemOpen
  \bibfield  {author} {\bibinfo {author} {\bibfnamefont {A.~K.}\ \bibnamefont
  {Naik}}, \bibinfo {author} {\bibfnamefont {M.~S.}\ \bibnamefont {Hanay}},
  \bibinfo {author} {\bibfnamefont {W.~K.}\ \bibnamefont {Hiebert}}, \bibinfo
  {author} {\bibfnamefont {X.~L.}\ \bibnamefont {Feng}}, \ and\ \bibinfo
  {author} {\bibfnamefont {M.~L.}\ \bibnamefont {Roukes}},\ }\href {\doibase
  10.1038/NNANO.2009.152} {\bibfield  {journal} {\bibinfo  {journal} {Nature
  Nanotechnology}\ }\textbf {\bibinfo {volume} {4}},\ \bibinfo {pages} {445}
  (\bibinfo {year} {2009})}\BibitemShut {NoStop}%
\bibitem [{\citenamefont {Rugar}\ \emph {et~al.}(2004)\citenamefont {Rugar},
  \citenamefont {Budakian}, \citenamefont {Mamin},\ and\ \citenamefont
  {Chui}}]{Rugar2004}%
  \BibitemOpen
  \bibfield  {author} {\bibinfo {author} {\bibfnamefont {D.}~\bibnamefont
  {Rugar}}, \bibinfo {author} {\bibfnamefont {R.}~\bibnamefont {Budakian}},
  \bibinfo {author} {\bibfnamefont {H.~J.}\ \bibnamefont {Mamin}}, \ and\
  \bibinfo {author} {\bibfnamefont {B.~W.}\ \bibnamefont {Chui}},\ }\href
  {\doibase 10.1038/nature02658} {\bibfield  {journal} {\bibinfo  {journal}
  {Nature}\ }\textbf {\bibinfo {volume} {430}},\ \bibinfo {pages} {329}
  (\bibinfo {year} {2004})}\BibitemShut {NoStop}%
\bibitem [{\citenamefont {Viola}\ \emph {et~al.}(1999)\citenamefont {Viola},
  \citenamefont {Knill},\ and\ \citenamefont {Lloyd}}]{Viola1999}%
  \BibitemOpen
  \bibfield  {author} {\bibinfo {author} {\bibfnamefont {L.}~\bibnamefont
  {Viola}}, \bibinfo {author} {\bibfnamefont {E.}~\bibnamefont {Knill}}, \ and\
  \bibinfo {author} {\bibfnamefont {S.}~\bibnamefont {Lloyd}},\ }\href@noop {}
  {\bibfield  {journal} {\bibinfo  {journal} {Physical Review Letters}\
  }\textbf {\bibinfo {volume} {82}},\ \bibinfo {pages} {2417} (\bibinfo {year}
  {1999})}\BibitemShut {NoStop}%
\bibitem [{\citenamefont {de~Lange}\ \emph {et~al.}(2011)\citenamefont
  {de~Lange}, \citenamefont {Rist\`{e}}, \citenamefont {Dobrovitski},\ and\
  \citenamefont {Hanson}}]{DeLange2011}%
  \BibitemOpen
  \bibfield  {author} {\bibinfo {author} {\bibfnamefont {G.}~\bibnamefont
  {de~Lange}}, \bibinfo {author} {\bibfnamefont {D.}~\bibnamefont {Rist\`{e}}},
  \bibinfo {author} {\bibfnamefont {V.~V.}\ \bibnamefont {Dobrovitski}}, \ and\
  \bibinfo {author} {\bibfnamefont {R.}~\bibnamefont {Hanson}},\ }\href
  {\doibase 10.1103/PhysRevLett.106.080802} {\bibfield  {journal} {\bibinfo
  {journal} {Physical Review Letters}\ }\textbf {\bibinfo {volume} {106}},\
  \bibinfo {pages} {080802} (\bibinfo {year} {2011})}\BibitemShut {NoStop}%
\bibitem [{\citenamefont {Sidles}\ \emph {et~al.}(1995)\citenamefont {Sidles},
  \citenamefont {Garbini}, \citenamefont {Bruland}, \citenamefont {Rugar},
  \citenamefont {Z\"{u}ger}, \citenamefont {Hoen},\ and\ \citenamefont
  {Yannoni}}]{Garbini1995}%
  \BibitemOpen
  \bibfield  {author} {\bibinfo {author} {\bibfnamefont {J.}~\bibnamefont
  {Sidles}}, \bibinfo {author} {\bibfnamefont {J.}~\bibnamefont {Garbini}},
  \bibinfo {author} {\bibfnamefont {K.}~\bibnamefont {Bruland}}, \bibinfo
  {author} {\bibfnamefont {D.}~\bibnamefont {Rugar}}, \bibinfo {author}
  {\bibfnamefont {O.}~\bibnamefont {Z\"{u}ger}}, \bibinfo {author}
  {\bibfnamefont {S.}~\bibnamefont {Hoen}}, \ and\ \bibinfo {author}
  {\bibfnamefont {C.}~\bibnamefont {Yannoni}},\ }\href {\doibase
  10.1103/RevModPhys.67.249} {\bibfield  {journal} {\bibinfo  {journal}
  {Reviews of Modern Physics}\ }\textbf {\bibinfo {volume} {67}},\ \bibinfo
  {pages} {249} (\bibinfo {year} {1995})}\BibitemShut {NoStop}%
\bibitem [{\citenamefont {Gruber}\ \emph {et~al.}(1997)\citenamefont {Gruber},
  \citenamefont {Drabenstedt}, \citenamefont {Tietz}, \citenamefont {Fleury},
  \citenamefont {Wrachtrup},\ and\ \citenamefont {von
  Borczyskowski}}]{Gruber1997}%
  \BibitemOpen
  \bibfield  {author} {\bibinfo {author} {\bibfnamefont {A.}~\bibnamefont
  {Gruber}}, \bibinfo {author} {\bibfnamefont {A.}~\bibnamefont {Drabenstedt}},
  \bibinfo {author} {\bibfnamefont {C.}~\bibnamefont {Tietz}}, \bibinfo
  {author} {\bibfnamefont {L.}~\bibnamefont {Fleury}}, \bibinfo {author}
  {\bibfnamefont {J.}~\bibnamefont {Wrachtrup}}, \ and\ \bibinfo {author}
  {\bibfnamefont {C.}~\bibnamefont {von Borczyskowski}},\ }\href@noop {}
  {\bibfield  {journal} {\bibinfo  {journal} {Science}\ }\textbf {\bibinfo
  {volume} {276}},\ \bibinfo {pages} {2012} (\bibinfo {year}
  {1997})}\BibitemShut {NoStop}%
\bibitem [{\citenamefont {Wrachtrup}\ and\ \citenamefont
  {Jelezko}(2006)}]{Wrachtrup2006}%
  \BibitemOpen
  \bibfield  {author} {\bibinfo {author} {\bibfnamefont {J.}~\bibnamefont
  {Wrachtrup}}\ and\ \bibinfo {author} {\bibfnamefont {F.}~\bibnamefont
  {Jelezko}},\ }\href@noop {} {\bibfield  {journal} {\bibinfo  {journal}
  {Journal of Physics-condensed Matter}\ }\textbf {\bibinfo {volume} {18}},\
  \bibinfo {pages} {S807} (\bibinfo {year} {2006})}\BibitemShut {NoStop}%
\bibitem [{\citenamefont {Balasubramanian}\ \emph {et~al.}(2009)\citenamefont
  {Balasubramanian}, \citenamefont {Neumann}, \citenamefont {Twitchen},
  \citenamefont {Markham}, \citenamefont {Kolesov}, \citenamefont {Mizuochi},
  \citenamefont {Isoya}, \citenamefont {Achard}, \citenamefont {Beck},
  \citenamefont {Tissler}, \citenamefont {Jacques}, \citenamefont {Hemmer},
  \citenamefont {Jelezko},\ and\ \citenamefont
  {Wrachtrup}}]{Balasubramanian2009}%
  \BibitemOpen
  \bibfield  {author} {\bibinfo {author} {\bibfnamefont {G.}~\bibnamefont
  {Balasubramanian}}, \bibinfo {author} {\bibfnamefont {P.}~\bibnamefont
  {Neumann}}, \bibinfo {author} {\bibfnamefont {D.}~\bibnamefont {Twitchen}},
  \bibinfo {author} {\bibfnamefont {M.}~\bibnamefont {Markham}}, \bibinfo
  {author} {\bibfnamefont {R.}~\bibnamefont {Kolesov}}, \bibinfo {author}
  {\bibfnamefont {N.}~\bibnamefont {Mizuochi}}, \bibinfo {author}
  {\bibfnamefont {J.}~\bibnamefont {Isoya}}, \bibinfo {author} {\bibfnamefont
  {J.}~\bibnamefont {Achard}}, \bibinfo {author} {\bibfnamefont
  {J.}~\bibnamefont {Beck}}, \bibinfo {author} {\bibfnamefont {J.}~\bibnamefont
  {Tissler}}, \bibinfo {author} {\bibfnamefont {V.}~\bibnamefont {Jacques}},
  \bibinfo {author} {\bibfnamefont {P.~R.}\ \bibnamefont {Hemmer}}, \bibinfo
  {author} {\bibfnamefont {F.}~\bibnamefont {Jelezko}}, \ and\ \bibinfo
  {author} {\bibfnamefont {J.}~\bibnamefont {Wrachtrup}},\ }\href@noop {}
  {\bibfield  {journal} {\bibinfo  {journal} {Nature Materials}\ }\textbf
  {\bibinfo {volume} {8}},\ \bibinfo {pages} {383} (\bibinfo {year}
  {2009})}\BibitemShut {NoStop}%
\bibitem [{\citenamefont {Cleland}(2003)}]{Cleland2003}%
  \BibitemOpen
  \bibfield  {author} {\bibinfo {author} {\bibfnamefont {A.~N.}\ \bibnamefont
  {Cleland}},\ }\href@noop {} {\emph {\bibinfo {title} {{Foundations of
  Nanomechanics: From Solid-State Theory to Device Applications}}}}\ (\bibinfo
  {publisher} {Springer},\ \bibinfo {year} {2003})\BibitemShut {NoStop}%
\bibitem [{\citenamefont {Marquardt}\ and\ \citenamefont
  {Girvin}(2009)}]{Marquardt2009}%
  \BibitemOpen
  \bibfield  {author} {\bibinfo {author} {\bibfnamefont {F.}~\bibnamefont
  {Marquardt}}\ and\ \bibinfo {author} {\bibfnamefont {S.}~\bibnamefont
  {Girvin}},\ }\href {\doibase 10.1103/Physics.2.40} {\bibfield  {journal}
  {\bibinfo  {journal} {Physics}\ }\textbf {\bibinfo {volume} {2}},\ \bibinfo
  {pages} {40} (\bibinfo {year} {2009})}\BibitemShut {NoStop}%
\bibitem [{\citenamefont {Aspelmeyer}\ \emph {et~al.}(2013)\citenamefont
  {Aspelmeyer}, \citenamefont {Kippenberg},\ and\ \citenamefont
  {Marquardt}}]{Aspelmeyer2013}%
  \BibitemOpen
  \bibfield  {author} {\bibinfo {author} {\bibfnamefont {M.}~\bibnamefont
  {Aspelmeyer}}, \bibinfo {author} {\bibfnamefont {T.~J.}\ \bibnamefont
  {Kippenberg}}, \ and\ \bibinfo {author} {\bibfnamefont {F.}~\bibnamefont
  {Marquardt}},\ }\href@noop {} {\bibfield  {journal} {\bibinfo  {journal}
  {arXiv}\ } (\bibinfo {year} {2013})},\ \Eprint
  {http://arxiv.org/abs/1303.0733v1} {arXiv:1303.0733v1} \BibitemShut {NoStop}%
\bibitem [{\citenamefont {Romero-Isart}\ \emph {et~al.}(2010)\citenamefont
  {Romero-Isart}, \citenamefont {Juan}, \citenamefont {Quidant},\ and\
  \citenamefont {Cirac}}]{Romero-Isart2010}%
  \BibitemOpen
  \bibfield  {author} {\bibinfo {author} {\bibfnamefont {O.}~\bibnamefont
  {Romero-Isart}}, \bibinfo {author} {\bibfnamefont {M.~L.}\ \bibnamefont
  {Juan}}, \bibinfo {author} {\bibfnamefont {R.}~\bibnamefont {Quidant}}, \
  and\ \bibinfo {author} {\bibfnamefont {J.~I.}\ \bibnamefont {Cirac}},\ }\href
  {\doibase 10.1088/1367-2630/12/3/033015} {\bibfield  {journal} {\bibinfo
  {journal} {New Journal of Physics}\ }\textbf {\bibinfo {volume} {12}},\
  \bibinfo {pages} {033015} (\bibinfo {year} {2010})}\BibitemShut {NoStop}%
\bibitem [{\citenamefont {Chang}\ \emph {et~al.}(2010)\citenamefont {Chang},
  \citenamefont {Regal}, \citenamefont {Papp}, \citenamefont {Wilson},
  \citenamefont {Ye}, \citenamefont {Painter}, \citenamefont {Kimble},\ and\
  \citenamefont {Zoller}}]{Chang2010}%
  \BibitemOpen
  \bibfield  {author} {\bibinfo {author} {\bibfnamefont {D.~E.}\ \bibnamefont
  {Chang}}, \bibinfo {author} {\bibfnamefont {C.~a.}\ \bibnamefont {Regal}},
  \bibinfo {author} {\bibfnamefont {S.~B.}\ \bibnamefont {Papp}}, \bibinfo
  {author} {\bibfnamefont {D.~J.}\ \bibnamefont {Wilson}}, \bibinfo {author}
  {\bibfnamefont {J.}~\bibnamefont {Ye}}, \bibinfo {author} {\bibfnamefont
  {O.}~\bibnamefont {Painter}}, \bibinfo {author} {\bibfnamefont {H.~J.}\
  \bibnamefont {Kimble}}, \ and\ \bibinfo {author} {\bibfnamefont
  {P.}~\bibnamefont {Zoller}},\ }\href {\doibase 10.1073/pnas.0912969107}
  {\bibfield  {journal} {\bibinfo  {journal} {Proceedings of the National
  Academy of Sciences of the United States of America}\ }\textbf {\bibinfo
  {volume} {107}},\ \bibinfo {pages} {1005} (\bibinfo {year}
  {2010})}\BibitemShut {NoStop}%
\bibitem [{\citenamefont {Li}\ \emph {et~al.}(2010)\citenamefont {Li},
  \citenamefont {Kheifets}, \citenamefont {Medellin},\ and\ \citenamefont
  {Raizen}}]{Li2010a}%
  \BibitemOpen
  \bibfield  {author} {\bibinfo {author} {\bibfnamefont {T.}~\bibnamefont
  {Li}}, \bibinfo {author} {\bibfnamefont {S.}~\bibnamefont {Kheifets}},
  \bibinfo {author} {\bibfnamefont {D.}~\bibnamefont {Medellin}}, \ and\
  \bibinfo {author} {\bibfnamefont {M.~G.}\ \bibnamefont {Raizen}},\ }\href
  {\doibase 10.1126/science.1189403} {\bibfield  {journal} {\bibinfo  {journal}
  {Science}\ }\textbf {\bibinfo {volume} {328}},\ \bibinfo {pages} {1673}
  (\bibinfo {year} {2010})}\BibitemShut {NoStop}%
\bibitem [{\citenamefont {Li}\ \emph {et~al.}(2011)\citenamefont {Li},
  \citenamefont {Kheifets},\ and\ \citenamefont {Raizen}}]{Li2011}%
  \BibitemOpen
  \bibfield  {author} {\bibinfo {author} {\bibfnamefont {T.}~\bibnamefont
  {Li}}, \bibinfo {author} {\bibfnamefont {S.}~\bibnamefont {Kheifets}}, \ and\
  \bibinfo {author} {\bibfnamefont {M.~G.}\ \bibnamefont {Raizen}},\ }\href
  {\doibase 10.1038/nphys1952} {\bibfield  {journal} {\bibinfo  {journal}
  {Nature Physics}\ }\textbf {\bibinfo {volume} {7}},\ \bibinfo {pages} {527}
  (\bibinfo {year} {2011})}\BibitemShut {NoStop}%
\bibitem [{\citenamefont {Yin}\ \emph {et~al.}(2013{\natexlab{a}})\citenamefont
  {Yin}, \citenamefont {Geraci},\ and\ \citenamefont {Li}}]{Yin2013a}%
  \BibitemOpen
  \bibfield  {author} {\bibinfo {author} {\bibfnamefont {Z.-Q.}\ \bibnamefont
  {Yin}}, \bibinfo {author} {\bibfnamefont {A.~a.}\ \bibnamefont {Geraci}}, \
  and\ \bibinfo {author} {\bibfnamefont {T.}~\bibnamefont {Li}},\ }\href
  {\doibase 10.1142/S0217979213300181} {\bibfield  {journal} {\bibinfo
  {journal} {International Journal of Modern Physics B}\ }\textbf {\bibinfo
  {volume} {27}},\ \bibinfo {pages} {1330018} (\bibinfo {year}
  {2013}{\natexlab{a}})}\BibitemShut {NoStop}%
\bibitem [{\citenamefont {Geraci}\ \emph {et~al.}(2010)\citenamefont {Geraci},
  \citenamefont {Papp},\ and\ \citenamefont {Kitching}}]{Geraci2010}%
  \BibitemOpen
  \bibfield  {author} {\bibinfo {author} {\bibfnamefont {A.~a.}\ \bibnamefont
  {Geraci}}, \bibinfo {author} {\bibfnamefont {S.~B.}\ \bibnamefont {Papp}}, \
  and\ \bibinfo {author} {\bibfnamefont {J.}~\bibnamefont {Kitching}},\ }\href
  {\doibase 10.1103/PhysRevLett.105.101101} {\bibfield  {journal} {\bibinfo
  {journal} {Physical Review Letters}\ }\textbf {\bibinfo {volume} {105}},\
  \bibinfo {pages} {101101} (\bibinfo {year} {2010})}\BibitemShut {NoStop}%
\bibitem [{\citenamefont {Arvanitaki}\ and\ \citenamefont
  {Geraci}(2013)}]{Arvanitaki2013}%
  \BibitemOpen
  \bibfield  {author} {\bibinfo {author} {\bibfnamefont {A.}~\bibnamefont
  {Arvanitaki}}\ and\ \bibinfo {author} {\bibfnamefont {A.~a.}\ \bibnamefont
  {Geraci}},\ }\href {\doibase 10.1103/PhysRevLett.110.071105} {\bibfield
  {journal} {\bibinfo  {journal} {Physical Review Letters}\ }\textbf {\bibinfo
  {volume} {110}},\ \bibinfo {pages} {071105} (\bibinfo {year}
  {2013})}\BibitemShut {NoStop}%
\bibitem [{\citenamefont {Yin}\ \emph {et~al.}(2011)\citenamefont {Yin},
  \citenamefont {Li},\ and\ \citenamefont {Feng}}]{Yin2011}%
  \BibitemOpen
  \bibfield  {author} {\bibinfo {author} {\bibfnamefont {Z.-Q.}\ \bibnamefont
  {Yin}}, \bibinfo {author} {\bibfnamefont {T.}~\bibnamefont {Li}}, \ and\
  \bibinfo {author} {\bibfnamefont {M.}~\bibnamefont {Feng}},\ }\href {\doibase
  10.1103/PhysRevA.83.013816} {\bibfield  {journal} {\bibinfo  {journal}
  {Physical Review A}\ }\textbf {\bibinfo {volume} {83}},\ \bibinfo {pages}
  {013816} (\bibinfo {year} {2011})}\BibitemShut {NoStop}%
\bibitem [{\citenamefont {Neukirch}\ \emph {et~al.}(2013)\citenamefont
  {Neukirch}, \citenamefont {Gieseler}, \citenamefont {Quidant}, \citenamefont
  {Novotny},\ and\ \citenamefont {{Nick Vamivakas}}}]{Neukirch2013}%
  \BibitemOpen
  \bibfield  {author} {\bibinfo {author} {\bibfnamefont {L.~P.}\ \bibnamefont
  {Neukirch}}, \bibinfo {author} {\bibfnamefont {J.}~\bibnamefont {Gieseler}},
  \bibinfo {author} {\bibfnamefont {R.}~\bibnamefont {Quidant}}, \bibinfo
  {author} {\bibfnamefont {L.}~\bibnamefont {Novotny}}, \ and\ \bibinfo
  {author} {\bibfnamefont {A.}~\bibnamefont {{Nick Vamivakas}}},\ }\href
  {\doibase 10.1364/OL.38.002976} {\bibfield  {journal} {\bibinfo  {journal}
  {Optics Letters}\ }\textbf {\bibinfo {volume} {38}},\ \bibinfo {pages} {2976}
  (\bibinfo {year} {2013})}\BibitemShut {NoStop}%
\bibitem [{\citenamefont {Scala}\ \emph {et~al.}(2013)\citenamefont {Scala},
  \citenamefont {Kim}, \citenamefont {Morley}, \citenamefont {Barker},\ and\
  \citenamefont {Bose}}]{Scala2013}%
  \BibitemOpen
  \bibfield  {author} {\bibinfo {author} {\bibfnamefont {M.}~\bibnamefont
  {Scala}}, \bibinfo {author} {\bibfnamefont {M.~S.}\ \bibnamefont {Kim}},
  \bibinfo {author} {\bibfnamefont {G.~W.}\ \bibnamefont {Morley}}, \bibinfo
  {author} {\bibfnamefont {P.~F.}\ \bibnamefont {Barker}}, \ and\ \bibinfo
  {author} {\bibfnamefont {S.}~\bibnamefont {Bose}},\ }\href {\doibase
  10.1103/PhysRevLett.111.180403} {\bibfield  {journal} {\bibinfo  {journal}
  {Physical Review Letters}\ }\textbf {\bibinfo {volume} {111}},\ \bibinfo
  {pages} {180403} (\bibinfo {year} {2013})}\BibitemShut {NoStop}%
\bibitem [{\citenamefont {Yin}\ \emph {et~al.}(2013{\natexlab{b}})\citenamefont
  {Yin}, \citenamefont {Li}, \citenamefont {Zhang},\ and\ \citenamefont
  {Duan}}]{Yin2013}%
  \BibitemOpen
  \bibfield  {author} {\bibinfo {author} {\bibfnamefont {Z.-Q.}\ \bibnamefont
  {Yin}}, \bibinfo {author} {\bibfnamefont {T.}~\bibnamefont {Li}}, \bibinfo
  {author} {\bibfnamefont {X.}~\bibnamefont {Zhang}}, \ and\ \bibinfo {author}
  {\bibfnamefont {L.~M.}\ \bibnamefont {Duan}},\ }\href {\doibase
  10.1103/PhysRevA.88.033614} {\bibfield  {journal} {\bibinfo  {journal}
  {Physical Review A}\ }\textbf {\bibinfo {volume} {88}},\ \bibinfo {pages}
  {033614} (\bibinfo {year} {2013}{\natexlab{b}})}\BibitemShut {NoStop}%
\bibitem [{\citenamefont {Kolkowitz}\ \emph
  {et~al.}(2012{\natexlab{b}})\citenamefont {Kolkowitz}, \citenamefont
  {{Bleszynski Jayich}}, \citenamefont {Unterreithmeier}, \citenamefont
  {Bennett}, \citenamefont {Rabl}, \citenamefont {Harris},\ and\ \citenamefont
  {Lukin}}]{Kolkowitz2012}%
  \BibitemOpen
  \bibfield  {author} {\bibinfo {author} {\bibfnamefont {S.}~\bibnamefont
  {Kolkowitz}}, \bibinfo {author} {\bibfnamefont {a.~C.}\ \bibnamefont
  {{Bleszynski Jayich}}}, \bibinfo {author} {\bibfnamefont {Q.}~\bibnamefont
  {Unterreithmeier}}, \bibinfo {author} {\bibfnamefont {S.~D.}\ \bibnamefont
  {Bennett}}, \bibinfo {author} {\bibfnamefont {P.}~\bibnamefont {Rabl}},
  \bibinfo {author} {\bibfnamefont {J.~G.~E.}\ \bibnamefont {Harris}}, \ and\
  \bibinfo {author} {\bibfnamefont {M.~D.}\ \bibnamefont {Lukin}},\ }\href
  {\doibase 10.1126/science.1216821} {\bibfield  {journal} {\bibinfo  {journal}
  {Science}\ }\textbf {\bibinfo {volume} {335}},\ \bibinfo {pages} {1603}
  (\bibinfo {year} {2012}{\natexlab{b}})}\BibitemShut {NoStop}%
\bibitem [{\citenamefont {Yang}\ \emph {et~al.}(2010)\citenamefont {Yang},
  \citenamefont {Wang},\ and\ \citenamefont {Liu}}]{Yang2010c}%
  \BibitemOpen
  \bibfield  {author} {\bibinfo {author} {\bibfnamefont {W.}~\bibnamefont
  {Yang}}, \bibinfo {author} {\bibfnamefont {Z.-Y.}\ \bibnamefont {Wang}}, \
  and\ \bibinfo {author} {\bibfnamefont {R.-B.}\ \bibnamefont {Liu}},\ }\href
  {\doibase 10.1007/s11467-010-0113-8} {\bibfield  {journal} {\bibinfo
  {journal} {Frontiers of Physics in China}\ }\textbf {\bibinfo {volume} {6}},\
  \bibinfo {pages} {1} (\bibinfo {year} {2010})}\BibitemShut {NoStop}%
\bibitem [{\citenamefont {Yang}\ and\ \citenamefont {Liu}(2008)}]{Yang2008}%
  \BibitemOpen
  \bibfield  {author} {\bibinfo {author} {\bibfnamefont {W.}~\bibnamefont
  {Yang}}\ and\ \bibinfo {author} {\bibfnamefont {R.-B.}\ \bibnamefont {Liu}},\
  }\href {\doibase 10.1103/PhysRevB.78.085315} {\bibfield  {journal} {\bibinfo
  {journal} {Physical Review B}\ }\textbf {\bibinfo {volume} {78}},\ \bibinfo
  {pages} {085315} (\bibinfo {year} {2008})}\BibitemShut {NoStop}%
\bibitem [{\citenamefont {Zhao}\ \emph {et~al.}(2011)\citenamefont {Zhao},
  \citenamefont {Hu}, \citenamefont {Ho}, \citenamefont {Wan},\ and\
  \citenamefont {Liu}}]{Zhao2011a}%
  \BibitemOpen
  \bibfield  {author} {\bibinfo {author} {\bibfnamefont {N.}~\bibnamefont
  {Zhao}}, \bibinfo {author} {\bibfnamefont {J.}~\bibnamefont {Hu}}, \bibinfo
  {author} {\bibfnamefont {S.}~\bibnamefont {Ho}}, \bibinfo {author}
  {\bibfnamefont {J.}~\bibnamefont {Wan}}, \ and\ \bibinfo {author}
  {\bibfnamefont {R.}~\bibnamefont {Liu}},\ }\href {\doibase
  10.1038/nnano.2011.22} {\bibfield  {journal} {\bibinfo  {journal} {Nature
  nanotechnology}\ }\textbf {\bibinfo {volume} {6}},\ \bibinfo {pages} {242}
  (\bibinfo {year} {2011})}\BibitemShut {NoStop}%
\bibitem [{\citenamefont {de~Lange}\ \emph {et~al.}(2010)\citenamefont
  {de~Lange}, \citenamefont {Wang}, \citenamefont {Riste}, \citenamefont
  {Dobrovitski},\ and\ \citenamefont {Hanson}}]{Lange2010}%
  \BibitemOpen
  \bibfield  {author} {\bibinfo {author} {\bibfnamefont {G.}~\bibnamefont
  {de~Lange}}, \bibinfo {author} {\bibfnamefont {Z.~H.}\ \bibnamefont {Wang}},
  \bibinfo {author} {\bibfnamefont {D.}~\bibnamefont {Riste}}, \bibinfo
  {author} {\bibfnamefont {V.~V.}\ \bibnamefont {Dobrovitski}}, \ and\ \bibinfo
  {author} {\bibfnamefont {R.}~\bibnamefont {Hanson}},\ }\href {\doibase
  10.1126/science.1192739} {\bibfield  {journal} {\bibinfo  {journal}
  {Science}\ }\textbf {\bibinfo {volume} {330}},\ \bibinfo {pages} {60}
  (\bibinfo {year} {2010})}\BibitemShut {NoStop}%
\bibitem [{\citenamefont {Takahashi}\ \emph {et~al.}(2008)\citenamefont
  {Takahashi}, \citenamefont {Hanson}, \citenamefont {van Tol}, \citenamefont
  {Sherwin},\ and\ \citenamefont {Awschalom}}]{Takahashi2008}%
  \BibitemOpen
  \bibfield  {author} {\bibinfo {author} {\bibfnamefont {S.}~\bibnamefont
  {Takahashi}}, \bibinfo {author} {\bibfnamefont {R.}~\bibnamefont {Hanson}},
  \bibinfo {author} {\bibfnamefont {J.}~\bibnamefont {van Tol}}, \bibinfo
  {author} {\bibfnamefont {M.~S.}\ \bibnamefont {Sherwin}}, \ and\ \bibinfo
  {author} {\bibfnamefont {D.~D.}\ \bibnamefont {Awschalom}},\ }\href@noop {}
  {\bibfield  {journal} {\bibinfo  {journal} {Physical Review Letters}\
  }\textbf {\bibinfo {volume} {101}},\ \bibinfo {pages} {47601} (\bibinfo
  {year} {2008})}\BibitemShut {NoStop}%
\bibitem [{\citenamefont {Bar-Gill}\ \emph {et~al.}(2012)\citenamefont
  {Bar-Gill}, \citenamefont {Pham}, \citenamefont {Belthangady}, \citenamefont
  {{Le Sage}}, \citenamefont {Cappellaro}, \citenamefont {Maze}, \citenamefont
  {Lukin}, \citenamefont {Yacoby},\ and\ \citenamefont
  {Walsworth}}]{Bar-Gill2011}%
  \BibitemOpen
  \bibfield  {author} {\bibinfo {author} {\bibfnamefont {N.}~\bibnamefont
  {Bar-Gill}}, \bibinfo {author} {\bibfnamefont {L.~M.}\ \bibnamefont {Pham}},
  \bibinfo {author} {\bibfnamefont {C.}~\bibnamefont {Belthangady}}, \bibinfo
  {author} {\bibfnamefont {D.}~\bibnamefont {{Le Sage}}}, \bibinfo {author}
  {\bibfnamefont {P.}~\bibnamefont {Cappellaro}}, \bibinfo {author}
  {\bibfnamefont {J.~R.}\ \bibnamefont {Maze}}, \bibinfo {author}
  {\bibfnamefont {M.~D.}\ \bibnamefont {Lukin}}, \bibinfo {author}
  {\bibfnamefont {A.}~\bibnamefont {Yacoby}}, \ and\ \bibinfo {author}
  {\bibfnamefont {R.}~\bibnamefont {Walsworth}},\ }\href {\doibase
  10.1038/ncomms1856} {\bibfield  {journal} {\bibinfo  {journal} {Nature
  communications}\ }\textbf {\bibinfo {volume} {3}},\ \bibinfo {pages} {858}
  (\bibinfo {year} {2012})},\ \Eprint {http://arxiv.org/abs/1112.0667}
  {arXiv:1112.0667} \BibitemShut {NoStop}%
\bibitem [{\citenamefont {Taylor}\ \emph {et~al.}(2008)\citenamefont {Taylor},
  \citenamefont {Cappellaro}, \citenamefont {Childress}, \citenamefont {Jiang},
  \citenamefont {Budker}, \citenamefont {Hemmer}, \citenamefont {Yacoby},
  \citenamefont {Walsworth},\ and\ \citenamefont {Lukin}}]{Taylor2008}%
  \BibitemOpen
  \bibfield  {author} {\bibinfo {author} {\bibfnamefont {J.~M.}\ \bibnamefont
  {Taylor}}, \bibinfo {author} {\bibfnamefont {P.}~\bibnamefont {Cappellaro}},
  \bibinfo {author} {\bibfnamefont {L.}~\bibnamefont {Childress}}, \bibinfo
  {author} {\bibfnamefont {L.}~\bibnamefont {Jiang}}, \bibinfo {author}
  {\bibfnamefont {D.}~\bibnamefont {Budker}}, \bibinfo {author} {\bibfnamefont
  {P.~R.}\ \bibnamefont {Hemmer}}, \bibinfo {author} {\bibfnamefont
  {A.}~\bibnamefont {Yacoby}}, \bibinfo {author} {\bibfnamefont
  {R.}~\bibnamefont {Walsworth}}, \ and\ \bibinfo {author} {\bibfnamefont
  {M.~D.}\ \bibnamefont {Lukin}},\ }\href@noop {} {\bibfield  {journal}
  {\bibinfo  {journal} {Nature Physics}\ }\textbf {\bibinfo {volume} {4}},\
  \bibinfo {pages} {810} (\bibinfo {year} {2008})}\BibitemShut {NoStop}%
\end{thebibliography}
\end{document}